\documentclass[namedreferences]{solarphysics}
\usepackage[hyperref,optionalrh,solaromanenum]{spr-sola-addons} 
\usepackage{graphicx}                    
\usepackage{amssymb}                     
\usepackage{color}                       
\usepackage{breakurl}                    
\usepackage{lscape}                      

\newcommand{\revone}{\bf \color{blue}}

\runningauthor{R.\ J.\ Leamon {\em et~al.}}
\runningtitle{Timing Terminators: Forecasting Sunspot Cycle 25 Onset}

\begin{document}

\begin{opening}

\title{Timing Terminators: Forecasting Sunspot Cycle 25 Onset}

\author[addressref={1,2},corref,email={robert.j.leamon@nasa.gov}]%
{\inits{R.J.}\fnm{Robert J.}~\lnm{Leamon}~\orcid{0000-0002-6811-5862}}
\author[addressref={3}]%
{\inits{S.W.}\fnm{Scott W.}~\lnm{McIntosh}~\orcid{0000-0002-7369-1776}}
\author[addressref={4}]{\inits{S.C.}\fnm{Sandra C.}~\lnm{Chapman}~\orcid{0000-0003-0053-1584}}
\author[addressref={4,5,6}]{\inits{N.W.}\fnm{Nicholas W.}~\lnm{Watkins}~\orcid{0000-0003-4484-6588}}

  \address[id={1}]{University of Maryland, Department of Astronomy, College Park, MD 20742, USA}
  \address[id={2}]{NASA Goddard Space Flight Center, Code 672, Greenbelt, MD 20771, USA.}
  \address[id={3}]{National Center for Atmospheric Research, High Altitude Observatory, P.O. Box 3000, Boulder, CO~80307, USA}
  \address[id={4}]{Centre for Fusion, Space and Astrophysics, University of Warwick, Coventry CV4 7AL, UK}
  \address[id={5}]{Centre for the Analysis of Time Series, London School of Economics and Political Science, London WC2A 2AZ, UK}
  \address[id={6}]{School of Engineering and Innovation, STEM Faculty, Open University, Milton Keynes, MK7 6AA, UK}
\begin{abstract}
Recent research has demonstrated the existence of a new type of solar event, the ``terminator.'' Unlike the Sun's signature events, flares and Coronal Mass Ejections, the terminator
{most likely originates} in the solar interior,
{at or near the tachocline}. 
The terminator signals the end of a magnetic activity cycle at the Sun's equator and the start of a sunspot cycle at mid latitudes. 
Observations indicate that the time difference between these events is very short, less than a solar rotation, in the context of the sunspot cycle. As the (definitive) start and end point of solar activity cycles the precise timing of terminators should permit new investigations into the meteorology of our star's atmosphere. In this letter we use a standard method in signal processing, the Hilbert transform, to identify a mathematically robust signature of terminators in sunspot records and in radiative proxies. 
Using 
{a linear extrapolation of the Hilbert phase of the sunspot number and F10.7 solar radio flux time series}
we can achieve higher fidelity 
{historical}
terminator timing than previous estimates have permitted. 
Further, this method presents a unique opportunity to project, 
{from analysis of sunspot data,}
when the next terminator will occur, 
{May 2020 ($+4$, $-1.5$ months),}
and trigger the growth of sunspot cycle 25. 
\end{abstract}

\end{opening}

\section{Introduction}
The quasi-decadal oscillation in the number of sunspots present on the Sun has been a driver of investigation since its discovery in 1844 \citep{1844AN.....21..233S} and became interchangeably known as the sunspot or solar cycle. Sixty years later it was noted that, as the number of spots swells to its maximum (at a time that became known as ``solar maximum'') and shrinks to its minimum number (at ``solar minimum'') over the course of 11(-ish) years, sunspots follow a migratory path from mid-latitudes (about $\pm35^\circ$) to their eventual disappearance a few degrees from the solar equator \citep{1904MNRAS..64..747M}. The spotless solar minimum period  ends abruptly when spots appear again at mid-latitudes and the long, slow, progression to the equator starts afresh. The abrupt appearance of sunspots defines the start of the next sunspot cycle and the latitudinal distribution of sunspots gives the appearance of butterfly wings. Since the middle of the last century the explanation (and prediction) of the 11(-ish) year variability and its partner ``butterfly diagram,'' as the heartbeat of solar activity, is one of the most prominent puzzles in solar physics \citep{1955ApJ...122..293P,1961ApJ...133..572B,1969ApJ...156....1L}. 

Over the last few years a new observational diagnostic technique has been applied to the understanding of solar variability \citep{2014ApJ...792...12M}. Ubiquitous small features observed in the Sun's extreme-ultraviolet corona, ``EUV Bright Points,'' or BPs \citep{1974ApJ...189L..93G, 2003ApJ...589.1062H, 2005SoPh..228..285M}, have been associated with tracing the evolution of the rotationally-driven giant convective scale \citep{2014ApJ...784L..32M} that had vertices that were dubbed ``g-nodes.'' Together, these features permit the tracking of the magnetic activity bands of the 22-year magnetic cycle of the Sun that extend the conventional picture of decadal-scale solar variability. Further, McIntosh and colleagues inferred that the global-scale (intra- and extra-hemispheric) interaction of these magnetic activity bands was required to explain the appearance and evolution of sunspots on the magnetic bands and thus to shape the solar cycle. 

The growth of new cycle sunspots follows a time when the low-latitude pair of oppositely polarized magnetic bands abruptly ``terminate'' at the equator \citep{2014ApJ...792...12M}. 
For example, the cycle 23 sunspots did not appear to grow in abundance or size until the cycle 22 bands had terminated (in late 1997). Similarly, the polarity mirror-image of this progression occurred in early 2011 for cycle 24 sunspots, following the termination of the cycle 23 bands. This equatorial termination, or cancellation, appears to signal the end of one sunspot cycle and leaves only the higher-latitude band in each hemisphere. Sunspots rapidly appear and grow on that mid-latitude band for several years in this, the ``ascending phase,'' until the next (oppositely-signed) band appears at high latitude. The presence of the new oppositely signed band triggers a downturn in sunspot production on the mid-latitude band; this occurrence defines the maximum activity level of that band and the start of a new extended cycle.
{\citet{2019NatSR...9.2035D} suggested that the most plausible 
mechanism for rapid transport of information 
from the equatorial termination of the old cycle's activity bands 
(of opposite polarity in opposite hemispheres)
to the mid-latitudes to trigger new-cycle growth
was a solar ``tsunami'' in the solar tachocline that migrates poleward with a gravity wave speed 
($\sim$300$\mbox{ km s}^{-1}$).}

\begin{figure}[p]
\centering
\includegraphics[width=0.9\linewidth]{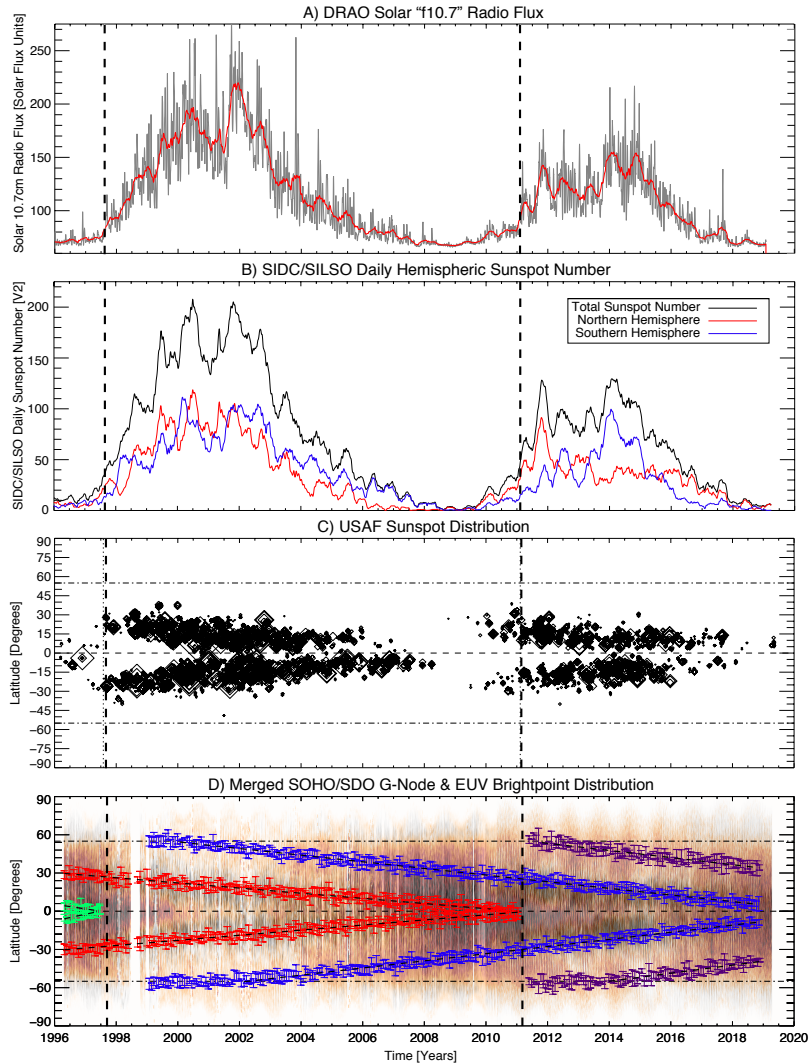}
\caption{Demonstrating the concept of Terminators and the brightpoint-activity band model and their relevance to the sunspot number during the {\em SOHO} epoch (1996\--2020). Panel A shows the daily (grey) and 50-day smoothed (red) 10.7cm solar radio flux from the Dominion Radio Astrophysical Observatory (DRAO). Panel B shows the daily (v2) hemispheric (red \-- north, blue \-- south) and total (black) sunspot numbers from Solar Influences Data Center (SIDC) of the Royal Observatory of Brussels. Each of the sunspot time series has a running 50-day smoothing. Panel C shows the United States Air Force (USAF) sunspot record---the size of the diamonds reflects the relative area of the sunspots in the record. Panel D shows the tracked centroids of the BP distribution for each hemispheric activity band, extending the work of \cite{2014ApJ...792...12M} -- cycle 22 bands in green, cycle 23 in red, 24 in blue and 25 in purple. The dashed vertical lines indicate the terminators \citep{McIntosh2019} of cycles~22 (August 1997) and~23 (February 2011). Extrapolating the fit for cycle 24 implies a termination date of April 2020 ($\pm 1$ month), and extrapolating cycle 25 predicts October 2031 $\pm 10$ months \citep{2018arXiv181202692L}.}
\label{fig:bands}
\end{figure}

Figure~\ref{fig:bands} illustrates the evolution of EUV BPs from 1996, at the minimum between sunspot cycles~22 and~23, to the present at the minimum between sunspot cycles~24 and~25 in context with the sunspot number, their latitudinal progression, and a signature measure of the Sun's radiative output---the 10.7cm solar radio flux. The large-scale magnetic activity bands that combine to shape sunspot cycles 22, 23, 24, and 25 are identified, as are the terminators. Note that, in both 1997\--98 and 2010\--11, the sunspot number has already started to increase from its activity minimum nadir since the bands temporally overlap. This is readily observed in comparison with panel C. Notice also the ``clumps'' of sunspots produced in each hemisphere and their corresponding signature in the total and hemispheric sunspot numbers \citep{2015NatCo...6.6491M, 2017NatAs...1E..86M}. Further, the terminators are clearly associated with a rapid increase in activity in (at least) one solar hemisphere \citep{2014ApJ...792...12M}.

In an effort to investigate sunspot cycle transitions and their terminators \cite{2014ApJ...792...12M}
used the 1997 and 2011 events as a guide. They (crudely) determined that a terminator had occurred when the total area of sunspots on the disk increased beyond the value of 100~millionths. This {\it ad hoc\/} definition was used to build a simplistic picture of magnetic activity band progression back over more than century.

\cite{McIntosh2019} 
returned to the topic of terminators
illustrating the presence of terminators in a range of standard solar diagnostics and sun-as-a-star activity proxies and discussing their relative importance in terms of understanding the solar interior. The data sets sampled spanned some 140 years of solar activity and illustrated a very abrupt event with a signature of enhanced magnetic flux emergence that leads to irradiance changes of a few percent in the lower atmosphere to almost 100\% in the corona over this short time frame---a veritable step function in activity. The Solar Cycle 23 to 24 transition is the best observed terminator to date (see, e.g., Fig.~\ref{fig:bands}), occurring in a few days around 11 February 2011, where observations from the twin STEREO and SDO spacecraft permitted a complete view of solar atmosphere. They also observed that the 2011 terminator was strongly longitudinal---the abrupt change at the equator and mid-latitudes could be observed with a distinct lag between the three spacecraft as they observed different solar longitudes---and the Sun transitioned from having one longitude of strong activity to five or six over the course of a few solar rotations following the terminator. 

In short, the analysis presented by \cite{McIntosh2019} reasonably validated the earlier, ad hoc, definition of a terminator, but the precise terminator timing was an issue to the reviewers of the work, especially as it concerned the data prior to 1996 and to when the next terminator may occur---the event that will trigger the growth of sunspot cycle 25. Those discussions motivate what follows---is there a robust (mathematical) signature of the terminator and when might sunspot cycle 25 spring forth?

We will use a standard method of signal processing---the Hilbert transform---applied to solar activity proxy time series ({\em i.e.}, the total and hemispheric sunspot number, and the 10.7cm solar radio flux) to investigate the accuracy of terminator timing and use this method to gain fidelity on when the next terminator will occur. 
Despite its utility in the signal-processing world, the application of the Hilbert transform to solar 
data
is remarkably limited: 
studies have focused either on very high frequency atmospheric fluctuations due to flare pulsations \citep{2015A&A...574A..53K},
or
investigating the long-term trends (or consistency) of the (envelope of the) 11-ish year sunspot cycle period  \citep{PhysRevLett.83.3406,2004AdSpR..34..302K,Barnhart2011,2016ApJ...830..140G}.

\section{Data and Methods}
\subsection{The Hilbert Transform}

In signal processing, the Hilbert transform is a specific linear operator that takes a function, $u(t)$ of a real variable and produces another function of a real variable ${\cal H}[u(t)]$ \citep{Bracewell2000,2002AmJPh..70..655P}. 
This linear operator is given by convolution with the function 
$1/(\pi t)$:
\begin{equation}
{\cal H}[u(t)]={\frac {\pm1}{\pi }}\int _{-\infty }^{\infty }{\frac {u(\tau)}{t-\tau }}\,d\tau ,
\label{eqn:1}
\end{equation}
the improper integral being understood in the Cauchy principal value sense. 
The Hilbert transform has a particularly simple representation in the frequency domain: it imparts a phase shift of $\pi/2$ to every Fourier component of a function; 
as such, an alternative interpretation is that the Hilbert transform is a ``differential'' operator, proportional to the time derivative of $u(t)$. 

A useful feature of the Hilbert transform becomes apparent by considering the complex time series $z(t)$ constructed from $u(t)$, now taken to have
(approximately)
zero mean, and its Hilbert transform $\mathcal{H}[u(t)]$ by
\begin{eqnarray}
    z(t) &=& u(t) + i {\cal H}[u(t)]  \\
         &=& A(t) \exp \left[ i \phi(t) \right] = A(t) \exp \left[ i \omega(t) t \right]. 
\label{eqn:4}
\end{eqnarray}

It is that analytic temporal phase $\phi(t)$ that we refer to above when referring to the Hilbert phase of SSN, F10.7, etc., variability.
{Noting that some authors choose to differ on the choice of $\pm$ sign in equation~(\ref{eqn:1}), 
we will adopt the $-1$ convention in 
defining the transform, such that $\phi$ decreases with time, 
a feature that more easily permits straightforward visual comparison with the decaying sunspot number timeseries; in either case, ${\cal H}[{\cal H}[u(t)]] = -u(t)$.
It also follows from equation~(\ref{eqn:4}) that $\omega = -d\phi / dt$, so slope of the changing phase with time has significance as a ``localized'' or ``instantaneous'' frequency of the fluctuating quantity \citep{Bracewell2000}.
}
A second useful feature of the Hilbert phase is in the phase coherence of two time series:
if edges/events in one time series occur at constant phase in another, the two are ``phase locked'' or ``synchronized'' \citep{2002AmJPh..70..655P,2013NatGe...6..289R,2018PhPl...25f2511C,2018NucFu..58l6003C}.

\begin{figure}[htp]
\centering
\includegraphics[width=0.95\linewidth]{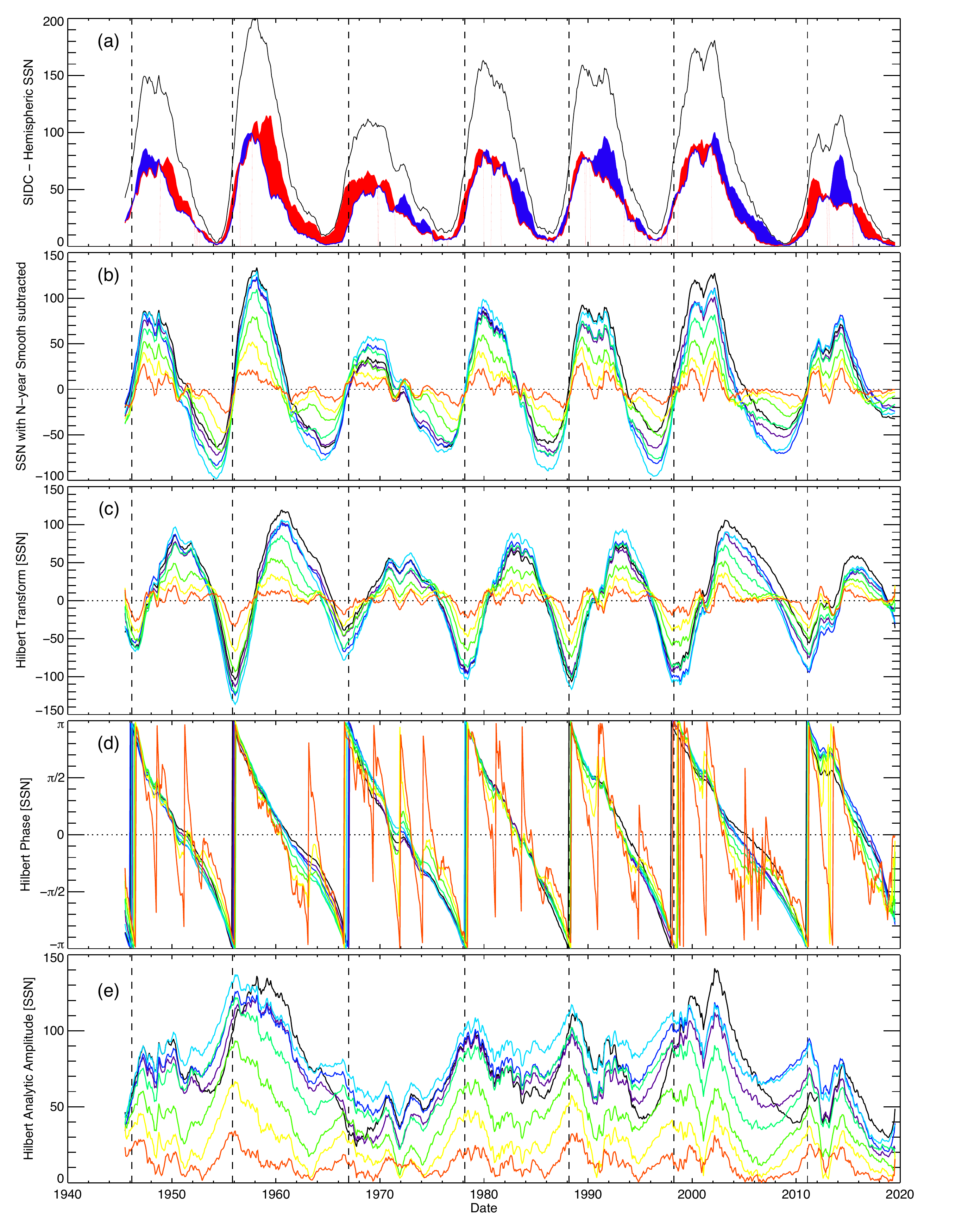}
\caption{Illustrating the properties of the Hilbert transform as applied to the sunspot record from 1947 to the present. 
(a) The total and hemispheric sunspot numbers, as recorded by the Royal Observatory of Belgium: the red and blue traces correspond to the northern and southern numbers respectively; colored fill corresponds to a dominance of the corresponding hemisphere over the other. 
(b) The total sunspot number from above, with an $N$-year running mean subtracted from it. The colored traces, from violet to red correspond to 30, 20, 15, 10, 7, 5, 3-year running means being subtracted before the Hilbert Transform computations, which we hold consistent throughout the Figure.
The black line corresponds to the constant mean $\langle R \rangle = 76.2$ of the whole time interval.
(c) The Hilbert transform, from equation~(\ref{eqn:1}), of each of the mean-subtracted SSN time series. The quarter-cycle phase shift is clear.
Panels
(d) and 
(e) show the Phase $\phi(t)$ and Amplitude $A(t)$, respectively, of the analytic signal from equation~(\ref{eqn:4}).
In panels~(b)--(e), In each panel the black dashed vertical lines correspond to the fitted crossings of the constant mean phase (black trace in panel~(d)) from $-\pi$ around to $\pi$ .
}
\label{fig:Havg}
\end{figure}

Figure~\ref{fig:Havg}
shows the foundation of our analyses and illustrates the essential patterns in the sunspot data and the properties of the Hilbert transform. 
The top panel shows the 
monthly hemispheric sunspot record from 1947 to the present as recorded by the Royal Observatory of Belgium. Although the blue and red fills correspond to variations of sunspots in one hemisphere or the other, 
we use as the example $u(t)$ their sum, the total sunspot number.
We use the IDL function {\tt hilbert.pro}\ which specifically uses the {\em discrete\/}  transform 
by computing a discrete Fourier transform, multiplying by $i$, and Fourier transforming back to the time domain. 

As mentioned above, the time series on which one computes the Hilbert transform should have a mean which is smaller than the excursions of the time-series, so that the phase around the unit circle monotonically in(de)creases and the frequency $\omega$ is positive on the timescale of physical interest.
Thus one may think of the input signal $u(t)$ being expressed as  
\begin{equation}
    u(t) = B(t) + A(t) \exp \left[ i \phi(t) \right]
\label{eqn:5}
\end{equation}
where $B(t)$ is that subtracted signal, the slow timescale trend, that is, slow compared to the frequency $\omega(t)$ of interest.
No information is created or destroyed, it is just book-kept in different places.

To demonstrate the effect of varying the ``slow'' timescale trend, 
the colored traces in panels (b)--(e) of Figure~\ref{fig:Havg}, from violet to red, correspond to 30, 20, 15, 10, 7, 5, 3-year running means being subtracted prior to further calculation, and the black line corresponds to 
$B(t)=$ constant $=\langle R \rangle$.
The robustness of the method is clear in that the same gross behavior is seen in all panels;
only in the 3- and 5-year (red and chartreuse) traces do we see deviations
and
more than one $-\pi$ around to $\pi$ phase crossing per solar cycle in panel~(d).

Figure~\ref{fig:Havg} shows that setting $B(t)$ to a constant (mean value) is sufficient to generate an analytical signal with monotonically increasing phase for the sunspot record over the last seven cycles. 
This will be the case provided that the peak amplitude of each solar cycle does not vary too violently between cycles. 
One could instead employ a time-varying slow timescale trend obtained using a more complex
local weight regression method such as LOWESS \citep{Cleveland1979.10481038} or a Savitsky-Golay filter \citep{SavGol64}; however these will inevitably suffer from edge effects. 
Since we will extrapolate the analytic phase beyond the edge to make a prediction for the next cycle, the optimal choice here is to remove a constant trend. 

In summary then, 
we can see from Figure~\ref{fig:Havg} that the choice of a whole-record mean, $B(t)=$ constant, is sufficient for the Hilbert transform over this time interval.
Further, 
since we will make a projection into the next solar cycle we are therefore justified using the constant whole-record mean.
We shall also discuss the coherent signature of the crossings of the phase from $-\pi$ around to $\pi$ a year or two after minimum in panel~(d) more fully in the following sections.

\subsection{Outline}

In the following sections we will demonstrate the utility of this functional decomposition by investigating the instantaneous amplitude and phase functions for a number of solar activity proxy time series: the total sunspot number over the past 200 years, the hemispheric sunspot number and the 10.7cm radio flux over the past 75 years. In each we'll see a characteristic signature in the amplitude and phase of the Hilbert transform at the times attributed to terminators in the literature \citep{McIntosh2019}. 
Because of this, 
we choose to keep the most information in $\phi(t)$, and so take 
$B(t)$ in equation~(\ref{eqn:5})  above as the constant mean $\langle u(t) \rangle$.
Finally, we develop this signature as a means to provide greater accuracy on when the next terminator, the one that will trigger the growth of solar cycle 25, will happen.
{We acknowledge that since the Hilbert phase wrap is mathematically consistent with the time of maximum rate of change of the underlying quantity, one could construct other methods of terminator proxy calculation, based on derivatives of the sunspot number time series. 
However, as Fig.~\ref{fig:Havg}(d) shows, the Hilbert phase wrap is a more robust indicator, especially when the data is ``noisy,'' with a first or second derivative equal to zero every second or third data point.}

\begin{figure}[htp]
\centering
\includegraphics[width=\linewidth]{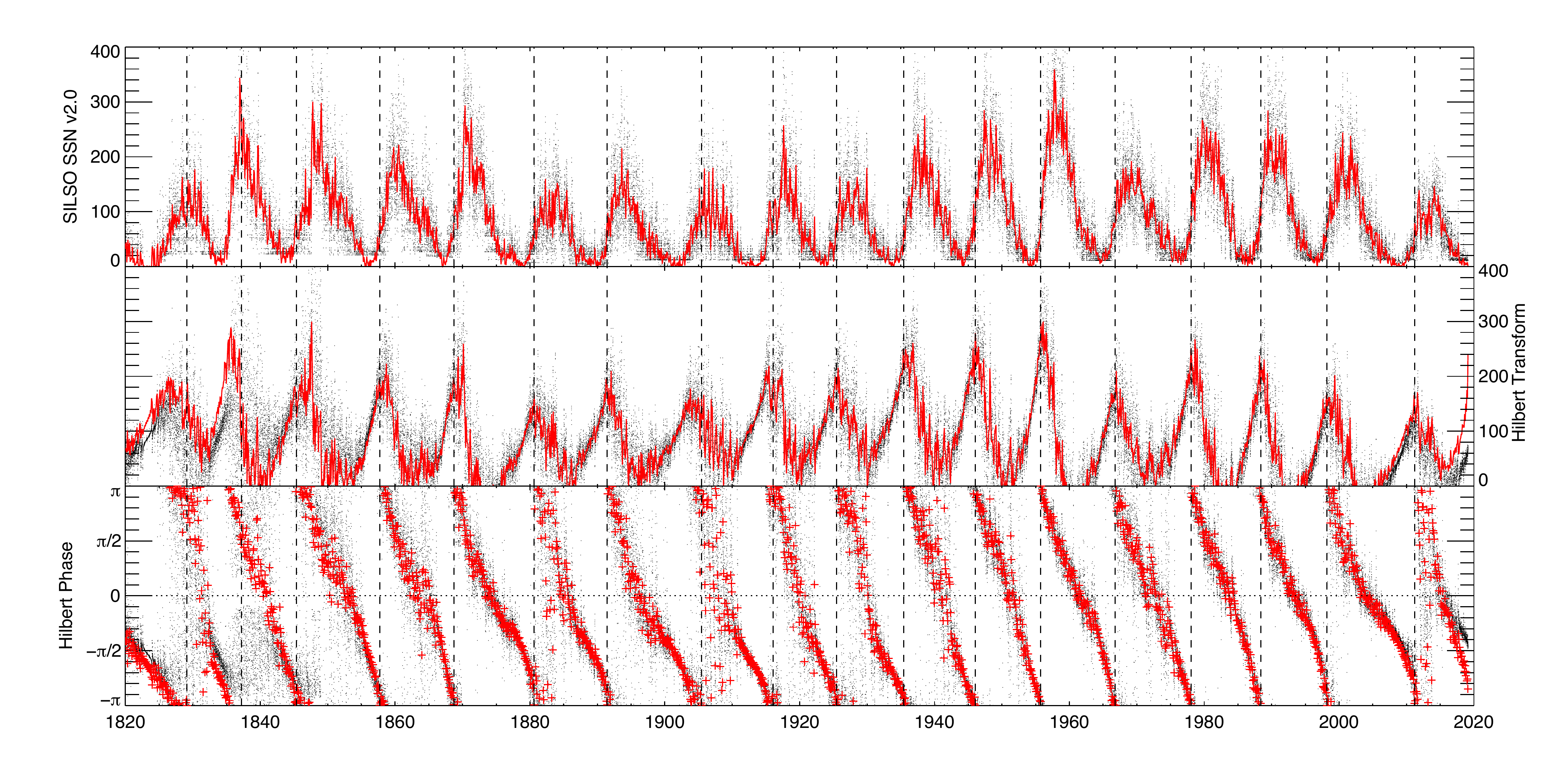}
\caption{Extending Fig.~\protect\ref{fig:Havg} to the total sunspot record, from 1820 to the present---Schwabe's 11-year cyclic behavior is clearly visible \citep{1844AN.....21..233S}. 
In the top panel we show the daily total sunspot number in black with the monthly sunspot number overplotted in red. In the center panel below we use these colors to illustrate the variation of the Hilbert transformed data for each time series. Similarly, in the bottom panel, we show the phase of the complex conjugate signal from equation~\ref{eqn:4}. In each panel the black dashed vertical lines correspond to the fitted crossings of the phase from $-\pi$ around to $\pi$.}
\label{fig:H2}
\end{figure}

\section{Results}

\subsection{200 Years of The Total Sunspot Number}

Figure~\ref{fig:H2} extends the analyses to daily  total sunspot number time series from 1812 to the Present.
The black and red time series in the middle panel represent the Hilbert transform of the zero-mean daily and monthly time series, respectively, with the mean added back in for ready comparison with the sunspot data above, 
again clearly
showing the $\pi/2$ phase shift between the two series. 
Similarly, the bottom panel shows the variation of the Hilbert Transform phase time series. Since $\sim$1845 (Cycle~9), we observe a striking pattern---the phase change from $-\pi$ to $\pi$ occurs a year or two after sunspot minimum and a year or two before sunspot maximum---around the point of maximum growth of the sunspot number. These points in time are marked on the plot by dashed vertical lines. Note that the phase of the Hilbert transform varies almost linearly from $\pi$ to $-\pi$ over the duration of the cycle although departures are clearly observed where there is a gradient change or ``knee,'' these will be discussed below.

Compared to the terminations shown in Fig.~\ref{fig:bands}, the Cycle~23 terminator in February 2011 is identical to the Hilbert transform computation; the Cycle~22 terminator in August 1997 is eight months ahead of the surge in the sunspot number (driven almost exclusively by the northern hemisphere). For the rest of this paper, we shall define the terminators to be represented by the date of the phase wrap of the whole-sun sunspot number as computed from Figure~\ref{fig:H2}.

\begin{figure}[htp]
\centering
\includegraphics[width=\linewidth]{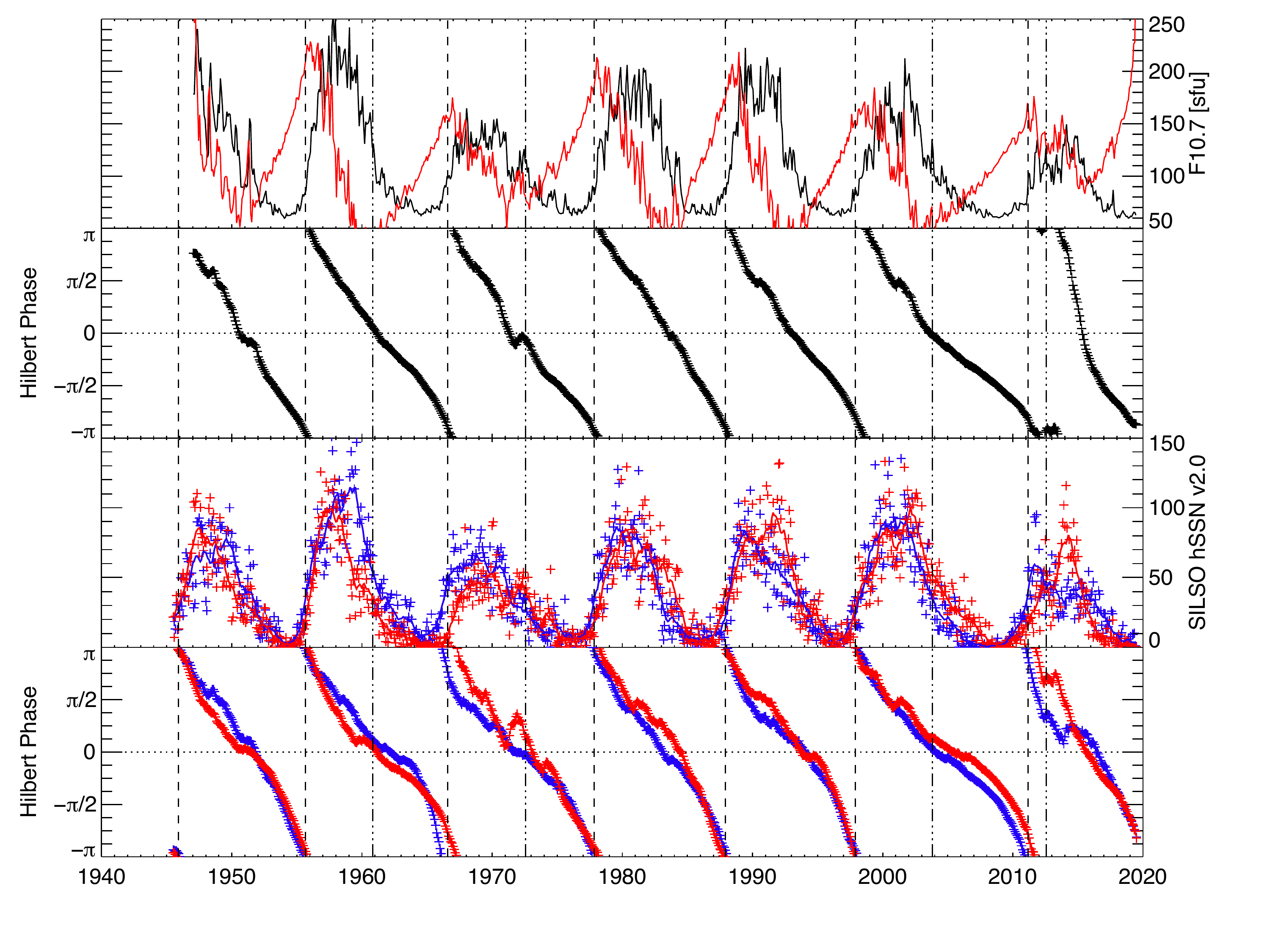}
\caption{Variation of the Hilbert transform amplitude and phase for the DRAO 10.7cm radio flux {\revone and} 
the hemispheric sunspot numbers over the past 75 years. The top panel show the variation of the 10.7cm radio flux timeseries (black) and its Hilbert transform (red) for comparison to the phase of the transform below. The lower set of panels show the hemispheric sunspot numbers (red \-- north; blue \-- south) and their phase. 
}
\label{fig:HF107}
\end{figure}

\subsection{75 Years of Radiative Proxies and the Hemispheric Sunspot Number}
We turn now to the application of the Hilbert transform to a widely used radiative proxy for solar activity, the 10.7cm radio flux (cf. Fig.~\ref{fig:bands}A) and the variation of the sunspot number in the Sun's two hemispheres. These records have been publically available for the last 75 years, or since just after the peak of solar cycle 18.

The upper panels of Fig.~\ref{fig:HF107} show that the amplitude and phase functions of the monthly averaged 10.7cm radio flux. In general, they exhibit the same properties as the total sunspot number with peaks in the former occurring after solar minimum but before solar maximum---at the strong step-like increases in coronal emission driven by the terminators \citep{Morgane1602056, 0004-637X-844-2-163,McIntosh2019}. As earlier, these are times of maximum change in the time series and correspond to the phase flips seen in the total sunspot number at the same time that are illustrated by dashed vertical lines.

The lower panels of Fig.~\ref{fig:HF107} show the application of this Hilbert transform method to the monthly hemispheric sunspot number. The monthly hemispheric sunspot numbers (blue for the northern hemisphere and red for the southern hemisphere) are shown as $+$ symbols and, for illustrative purposes, a 12-month running average is also shown as a relevantly colored solid line. The corresponding phase functions, for the monthly hSSN data, are shown in the lowest panel. 

Note that the characteristics of the amplitude and phase functions shown in Fig.~\ref{fig:HF107} mirror those of the total sunspot number in Fig.~\ref{fig:H2}. The 10.7cm radio flux shows and amplitude function maximum and a phase function that flips sign at times we have previously attributed to terminators. Interestingly, the phase functions resulting from the hemispheric sunspot numbers can separate by as much as a year when approaching the phase flip, but exhibit the same characteristic behavior of the total sunspot number, albeit with the expected, subtle, differences between the Sun's hemispheres \citep{2013ApJ...765..146M}.

Visible also in the phase function plots of Fig.~\ref{fig:HF107} are ``knees,'' or clear gradient changes in the phase function. Examples are visible in 1959, 1960, 1972, and 2003 and marked by dot-dashed lines. Note that those knees appear as a hemispheric signature and that the most pronounced, like that in 2003, are visible also in the phase function of the sun-as-a-star 10.7cm radio flux proxy, or the total sunspot number (Fig.~\ref{fig:H2}). We will return to a discussion of these phase function knees below.

\subsection{When will Cycle 24 Terminate?}
Identifying the start and end of solar cycles is a topic of some debate in solar community and estimates can range wildly \citep{Pesnell2008}. 
The identification of terminators as the trigger for the growth of mid-latitude sunspot formation changes that narrative, {as a single precise event, rather than, say, the nadir of solar minimum which is the overlap of the old and new cycles, in each of two hemispheres, and not subject to an annual-scale smoothing. 
In late 2019, at the time of this writing,} a timely question is when will solar cycle 25 start? In other words, when will the bands of cycle 24 terminate at the solar equator?

\cite{2014ApJ...792...12M} performed a linear extrapolation of the equatorward progress of the Cycle 24 activity bands (e.g., Figure~\ref{fig:bands}) visible in 2013 and anticipated that the terminator would arise in late 2019 or early 2020. In a subsequent paper, with updated data and band centroid tracking supported that conclusion \citep{2017FrASS...4....4M}. As a check on those predictions, we can use the phase functions of the hemispheric sunspot number of Figure~\ref{fig:HF107} as a guide and linearly extrapolate the roughly linear portion of the last few years (specifically, from {August 2017} on) to estimate when the phase function flip might occur.

\begin{figure}[tp]
\centering
\includegraphics[width=\linewidth]{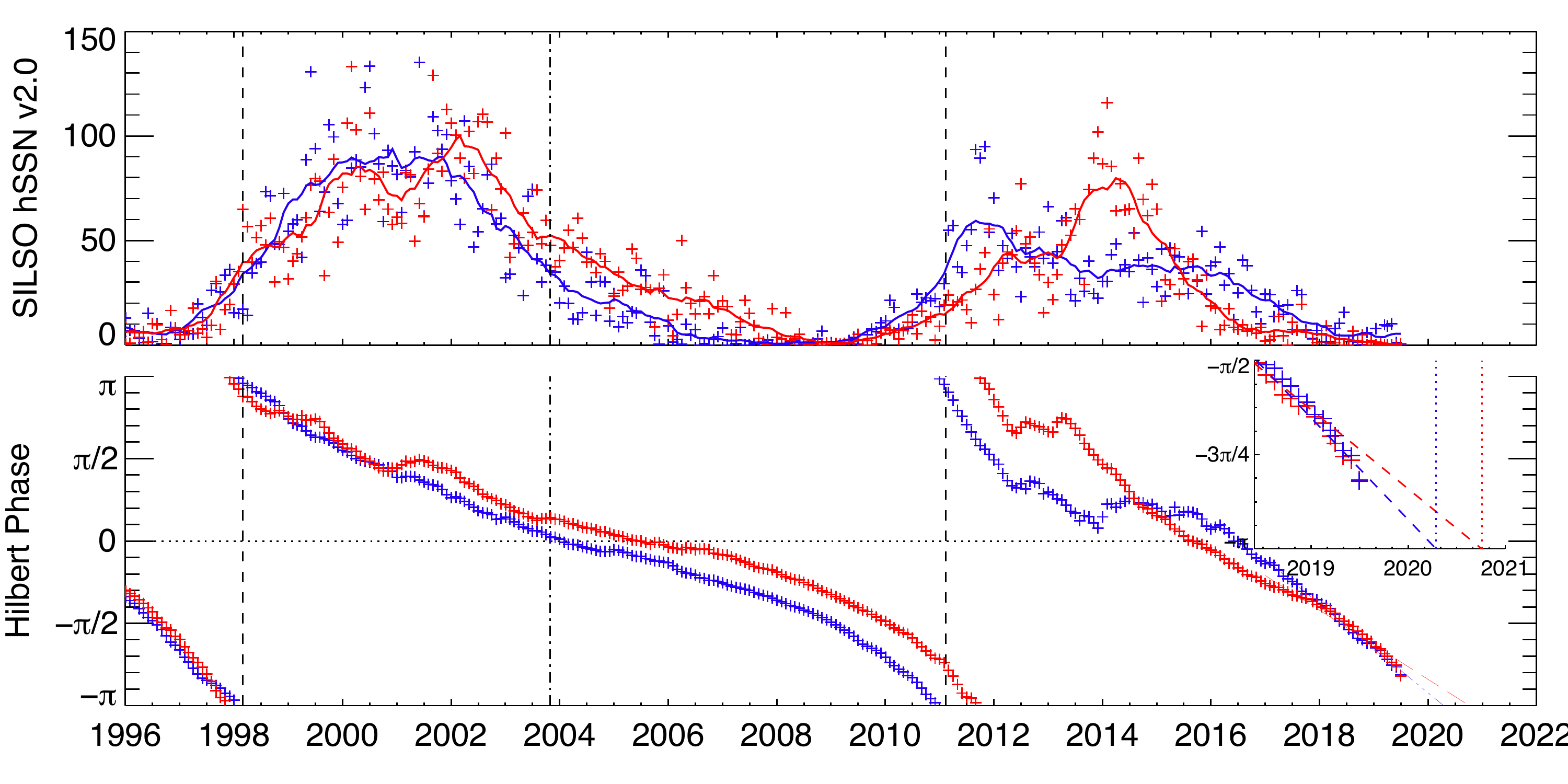}
\caption{The termination of sunspot cycle 24. Restricting the timeframe shown in Fig.~\ref{fig:H2} to the SOHO-era (1996--Present) we gain a little more fidelity on the phase functions of the hemispheric sunspot number (hSSN; top panel) where the monthly hSSN values (blue - north; red - south) are shown as $+$ symbols and their running 12 month average as a solid colored line. The corresponding phase functions, for the monthly hSSN data, are shown in the lower panel. In the upper right corner of this lower panel we show an linear extrapolation (dashed line) to the hemispheric phase functions using values for {August 2017} to present. Where these extrapolated lines cross the y-axis (a phase function value of $-\pi$) we have drawn vertical dashed lines. These points represent the outer limits of the anticipated phase-flip and hence the termination of the solar cycle 24 bands and rapid growth of sunspot cycle 25. These lines are correspond to March (north, $\pm 2$ mo) and September (south, $\pm 1$ mo) of 2020.}
\label{fig:Hh1996}
\end{figure}

Figure~\ref{fig:Hh1996} shows the hemispheric sunspot number over the SOHO epoch (1996--Present, as per Figure~\ref{fig:bands}) to show the transition from cycles 22 to 23 and 24 to 25. In the lower panel we show linear extrapolations of the hemispheric phase function variation from 2016 until the time of writing. 
The extrapolations of the northern and southern phase functions---shown in more detail in the inset of the lower panel---indicate that a phase flip will occur around March 2020 (North) and September 2020 (South). Further, inspection of the end-of-cycle behavior of the phase
({\em i.e.}, when linear behavior breaks down) 
in Figures~\ref{fig:H2} and~\ref{fig:HF107} implies a linear extrapolation can be late by as much as six months---placing the phase flip slightly earlier than April 2020. 
Note that the same linear analysis of the total sunspot number yields an approximate phase flip time of May~2020.
{Repeating the same analysis for the F10.7 data, 
we get February 2020 ($\pm 1$ mo) for the phase flip prediction.}

\begin{figure}[tp]
\centering
\includegraphics[width=\linewidth]{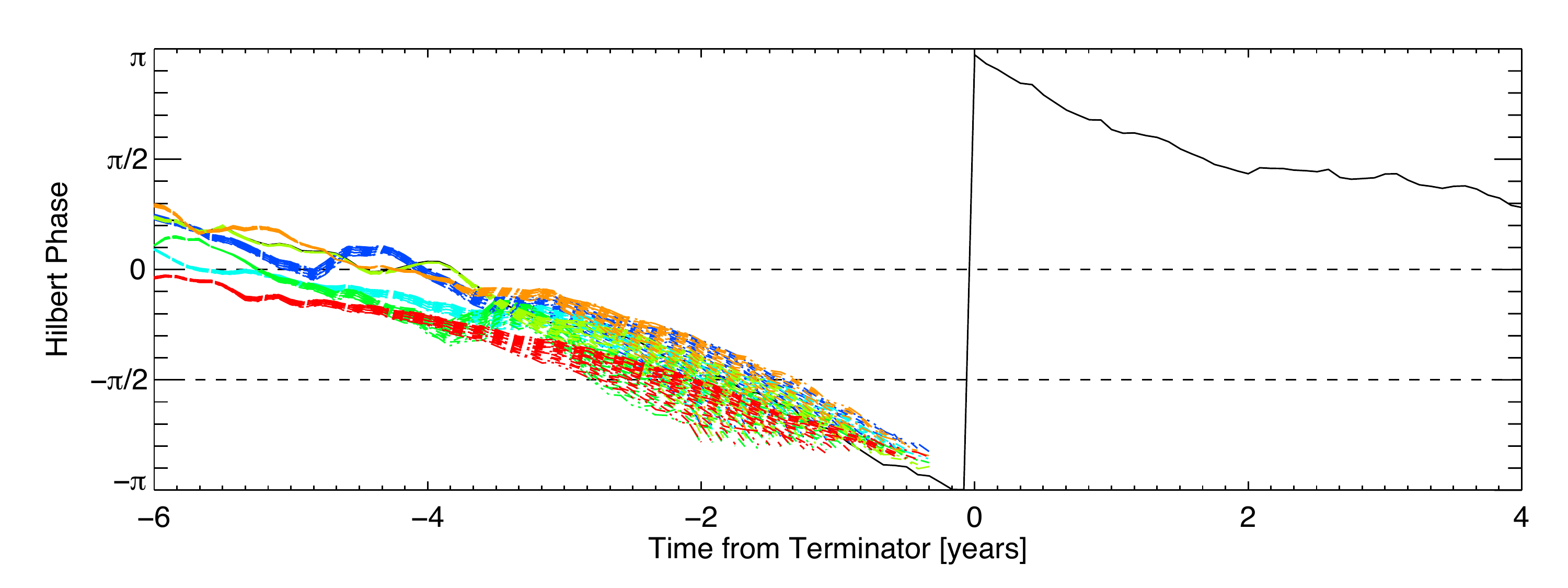}
\includegraphics[width=\linewidth]{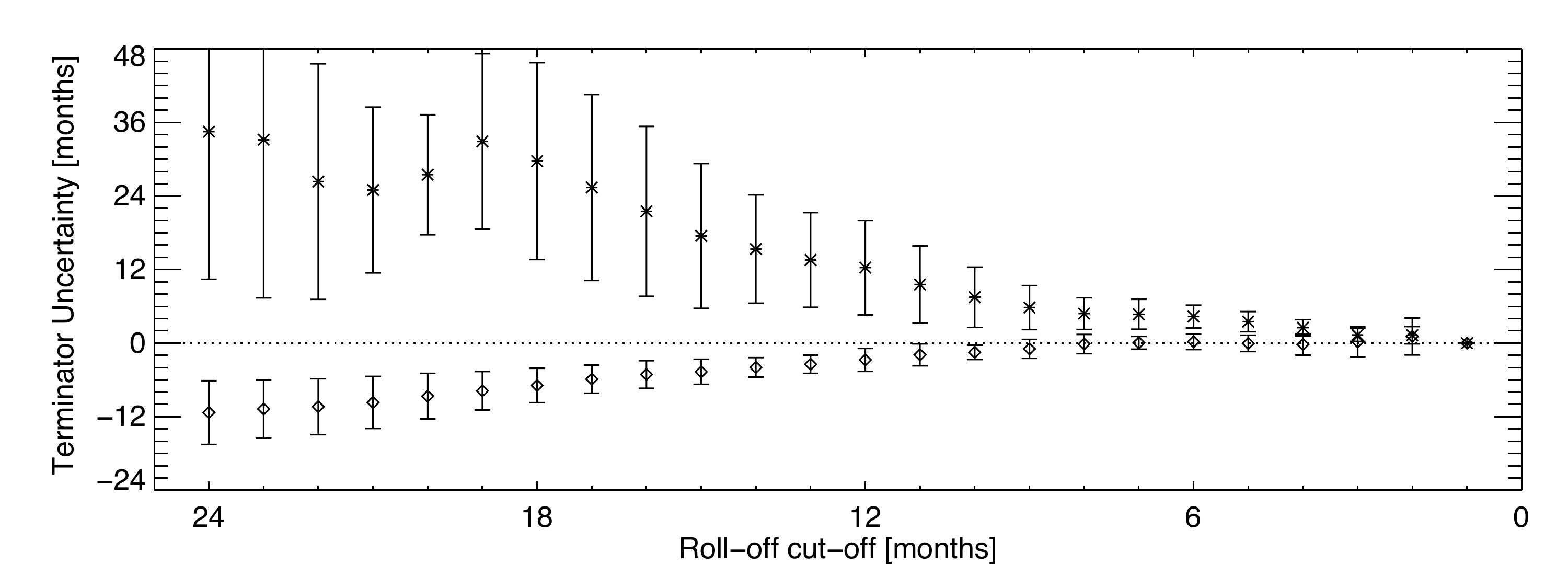}
\caption{{Investigating the edge effects of a shorter record on the ability of the Hilbert phase extrapolation method for predicting the Cycle~23 and~24 terminators, using the whole-Sun sunspot number record.
(top) Superposed epoch analysis of cycles 18-23 showing the edge effects of removing one extra data point (one month) from the record before computing the Hilbert transform. Each cycle has 20 traces showing the effects of removing 4, 5, \ldots, 24 data points.
{\revone The colors blue, cyan, green, chartreuse, orange and red correspond to Cycles 18-23 respectively; the underlying black trace shows the observed Cycle 21 data as reference.}
(bottom) The decreasing uncertainty of edge effects on prediction as one gets closer to the observed terminator date.
For each ordinate representing excluding $n$ months, 
the Diamonds represent the {\em earlier\/} prediction of the effects of Hilbert edge effect roll-off fitting the last 24 included data points; 
the 
Asterisks represent the {\em later\/} predictions of fitting the last 18 months prior to excluding $2n$ months, when rollover effects are minimised.
The error bars are the standard deviations of the six cycles.
}}
\label{fig:HRoll}
\end{figure}

\subsection{Quantifying Prediction Uncertainties}

One may query the relative tightness of the uncertainty bounds discussed in the above paragraph and in Figure~\ref{fig:Hh1996}: no worse than $\pm 2$ months.
The quoted uncertainty is propagated from the uncertainties in the linear fit over 24 points (August 2017 to September 2019). At the time of writing, we are confident in the tightness of that window.
However, we have the benefits of data through September 2019 to make that fit and prediction.
Could we have made that same prediction in August 2017, say?
There are well known and unavoidable edge effects with the Hilbert phase
determination \citep{2002AmJPh..70..655P}.
To investigate the effects of edge effects on predictions, we consider the top panel of Figure~\ref{fig:HRoll}, which follows on from the lower panel of Figure~\ref{fig:Hh1996}.
All curtailed records agree on the occurrence of past terminators, but the edge effects affect the prediction of the next terminator.
We see that (specifically for this method and this data), the edge effect is a roll-off on a timescale $R$ of approximately 9 months, and that curtailing the data by $n$ months leads to prediction about $n$ months earlier. 
Therefore any linear fitting that includes phases within $R=9$ months of the end of the record will always 
underestimate (get a time too early) for the next terminator.
However, conversely, looking at previous cycles in Figures~\ref{fig:Hh1996} and~\ref{fig:HRoll}, we can see that the phase can naturally show a roll-off (knee): so if we do not use phases within $R$ of the terminator we will tend to overestimate (get a time too late).
We can take these two predictions as the lower and upper bounds, respectively, of the uncertainty as a function of roll-off record truncation.
This is shown in the bottom panel of Figure~\ref{fig:HRoll}. The uncertainties decrease with a later fit interval, and asymptote towards the terminator.
Given that we are 5 months to the predicted date of Cycle 24 termination,
the uncertainty bound is better described by taking not the linear fit uncertainty, but the 5-month cut-off numbers from Figure~\ref{fig:HRoll}, thus
May 2020 (${}^{+4}_{-1.5}$ months).

{Finally, in} 
Figure~\ref{fig:Hh1996}, 
both the Cycle 22 minimum in 1996--97 and the Cycle 23 minimum in 2008--09 occur at phases $\sim -\pi/2$.
This phase occurred in mid-2018 for the present cycle, implying that Cycle~24 minimum {\em has already happened}. 
Even accounting for edge effects in the Hilbert and averaging process \citep{1999ITSP...47.2600M}, 
it is
highly unlikely that the current minimum will be any later than {mid-late} 2019, and predictions of another extended minimum will prove to be false.
This is borne out by torsional oscillation data
(Scherrer, 2019, personal communication\footnote{And his Hale Prize lecture, St.~Louis AAS SPD meeting.}) 
and the 
SIDC extrapolations using the
\cite{1949TrAGU..30..673M}
methodology.

\section{Discussion}
We have shown that the Hilbert Transform offers a means to develop a rigorous, mathematical description and identification of solar cycle termination points without access to complex datasets ({\em i.e.}, the distribution of BPs on the solar disk, the original definition). 
For the Sun, these epoch changes are not just some random point in time between solar minima and maxima---the terminators mark {\em the} start of periods of intense mid-latitude activity triggered by the death of the bands at the equator. Table~\ref{T1} provides the reader with a table of climatological magnetic/solar cycle times, including the times of hemispheric sunspot maxima \citep{2014ApJ...784L..32M}, terminators derived from the total and hemispheric sunspot numbers, the times between consecutive terminators, and other values. Note that the terminator values shown are both internally consistent and also not in vast disagreement with the values derived from the ad hoc sunspot area criterion \citep{2014ApJ...784L..32M}.

Extrapolation of the Hilbert phase functions of the hemispheric sunspot number indicate that the Termination of cycle 24 will occur in mid-2020, consistent with the earlier predictions from BP migration tracks alone \citep{2014ApJ...784L..32M, 2017FrASS...4....4M}, and the sunspots of cycle 25 should rapidly grow thereafter \citep{McIntosh2019}. 

One of the most interesting features in the Hilbert phase functions, beyond the terminator-related phase flip, are the knees. It is intriguing to contemplate that the knee of the phase function could be the result, and hence also a diagnostic, of significant eruptive activity. 
The knees mark the slowing down of the magnetic bands progression to the equator. 
Based on the phenomenological model of \cite{2014ApJ...792...12M}, slowing down the bands would increase the time of overlap between magnetic bands of two cycles and produce a longer, shallower, declining phase of the solar cycle.
If we associate the slope of the phase function with the ``rate of progression'' of the solar cycle, 
{then the knee or break in the phase function, clearly visible in late 2003 (see Figs.~\ref{fig:H2}, \ref{fig:HF107}  and \ref{fig:Hh1996}, and delineated with dot-dot-dashed vertical lines) implies that}
Cycle~23 dramatically slows down.
Since the inflection happens around $\phi = 0$, we can say that the first ``half'' of Cycle~23 lasted 5 years 2 months (from 1998 August to 2003 October), and the second phase with its unusual solar minimum \citep{doi:10.1063/1.3395955} lasted 7 years 4 months (to 2011 February). 
Note that this extended sunspot minimum also deceived many of the community experts that sat on the 2007 NOAA Solar Cycle 24 Forecast panel \citep{Pesnell2008}.

{October 2003, of course, is well known for the}
``Halloween storms,'' the series of powerful flares (17 total X-class, including an X29, and the estimated X47 flare on November 4, the largest flare recorded), primarily from Active Regions 10486 and 10488 and multiple Ground Level Enhancements \citep[GLEs]{PhysRev.104.768}---when the Sun emits particles of sufficient energy and intensity to raise radiation levels on Earth's surface. 
We wonder if large events such as the Halloween storms are at least partly responsible for longer than average terminator-to-terminator times seen in  Table~\ref{T1} (column $\Delta$). 
Further, if so, might the phase function knees be another possible means to investigate the occurrence of large historic solar eruptions before routine (detailed) observation? 
Were this scenario possible it would be a case of the ``tail wagging the dog,'' but 
it is not completely unfeasible, as the Halloween storms alone removed most (just over half) of the Solar Cycle~23 cumulative helicity budget \citep{2005JGRA..110.8107L}.

{ In fact,}
similar knees are readily seen in Figure~\ref{fig:HF107} in November 1960, on the downslope of cycle~19 (18 months after the Northern Hemisphere maximum, and almost 3 years after the overall cycle maximum). We speculate, then, that the three large flares and associated GLEs in three days from ``McMath Plage 5925'' at 29${}^\circ$N\footnote{The present consecutive Active Region numbering system only started in January 1972.}---unusually high for that phase of the solar cycle, so likely due to interaction with the following cycle---entirely consistent with the activity band model prediction \citep{2014ApJ...796L..19M,2017FrASS...4....4M}.
{ We acknowledge, however, 
that to avoid (the appearance of) selection bias, 
all major eruptions and knees need to be carefully considered in a future work, as well as the subtle differences between the SSN and F10.7 radio flux traces in Figure~\ref{fig:HF107}.
Nevertheless, speculation over the cause of the knees does not change the observed cycle lengths, nor the results of this work predicting when the current solar cycle will terminate.}

\section{Conclusion}
We have employed a standard method in signal processing, the Hilbert transform, to identify a mathematically robust signature of terminators in sunspot records and in radiative proxies. Using this technique we can achieve higher fidelity 
{historical} terminator timing than previous estimates have permitted. 
Further, this method presents a unique opportunity to project,
{from analysis of sunspot data,}
that the next terminator will soon occur, 
{May 2020 (${}^{+4}_{-1.5}$ months),}
and trigger the growth of sunspot cycle 25.

\vspace{2em}
This material is based upon work supported by the National Center for Atmospheric Research, which is a major facility sponsored by the National Science Foundation under Cooperative Agreement No.\ 1852977.
RJL acknowledges support from NASA's Living With a Star Program.

\begin{landscape}
\begin{table}[tp]
\caption{Revised version M2014's Table~1 with Terminator dates as calculated via the Hilbert Transform phase function, as shown in Figure~\protect\ref{fig:H2}. From left to right we provide the sunspot cycle number; the start times of the bands in each hemispheric (the time of the previous hemispheric maxima); the time between hemispheric maxima of the same magnetic polarity ($\delta_{N}$ and $\delta_{S}$); the terminator derived from the total sunspot number (\protect\ref{fig:H2}); the time elapsed since last terminator ($\Delta$); the terminator values determined from the hemispheric sunspot number (\protect\ref{fig:HF107}); and the hemispheric high–low latitude transit time ($\tau$) for Each Hemisphere. As a reminder, $T_{N,S}^{\it max}$ is hemispheric sunspot maximum and, as per M2014, is also the first incursion of the band that becomes the {\em upcoming\/} sunspot cycle.}
 \label{T1}
 \begin{center}
 \begin{tabular}{ccccc|cccc|ccc}
    \hline
	Cycle & $T_N^{\it max}$ (yr) & $T_S^{\it max}$ (yr) & 
    $\delta_{N}$ (yr) & $\delta_{S}$ (yr) & Term.$^{SSN}$ (yr) & 
    $\Delta$ (yr) & $T_N^t$ (yr) & $T_S^t$ (yr) & $\tau_{N}$ (yr) & $\tau_{S}$ (yr) \\
	\hline
    12 & $\ldots$ & $\ldots$ & $\ldots$ & $\ldots$ & 1891.75 & 9.58 & $\ldots$ & $\ldots$ & $\ldots$ & $\ldots$  \\    
    13 & 1884.00 & 1883.83 & $\ldots$ & $\ldots$ & 1905.00 & 13.25 & $\ldots$ & $\ldots$ & 20.75 & 20.92 \\
    14 & 1892.50 & 1893.58 & $\ldots$ & $\ldots$ & 1916.75 & 11.75 & $\ldots$ & $\ldots$ & 22.55 & 21.47 \\
    15 & 1905.75 & 1907.08 & 21.75 & 23.25 & 1925.50 & 8.75 & $\ldots$ & $\ldots$ & 19.92 & 18.59 \\
    16 & 1917.58 & 1919.50 & 25.08 & 25.92 & 1935.75 & 10.25 & $\ldots$ & $\ldots$ & 18.17 & 16.25 \\
	17 & 1925.92 & 1926.08 & 20.17 & 19.00 & 1945.75 & 10.00 & 1946.42 & 1946.33 & 19.83 & 19.67 \\
    18 & 1937.50 & 1939.67 & 19.92 & 20.17 & 1955.67 & 9.92 & 1955.83 & 1956.16 & 18.25 & 16.08 \\
    19 & 1949.17 & 1947.17 & 23.25 & 21.08 & 1966.67 & 11.00 & 1966.58 & 1967.58 & 17.33 & 19.33 \\
	20 & 1959.08 & 1956.83 & 21.58 & 17.17 & 1977.92 & 11.25 & 1978.08 & 1978.50 & 18.92 & 21.17 \\
    21 & 1968.00 & 1970.08 & 18.83 & 22.92 & 1988.08 & 10.16 & 1988.33 & 1988.25 & 20.50 & 18.42 \\
    22 & 1979.67 & 1980.33 & 20.58 & 23.50 & 1998.67 &  10.58 & 1998.50 & 1998.25 & 18.08 & 17.42 \\
	23 & 1989.08 & 1991.08 & 21.08 & 21.00 & 2011.17 & 12.50 & 2011.25 & 2012.16 & 22.12 & 20.12 \\
    24 & 2000.50 & 2002.58 & 20.83 & 22.25 & [2020.33] & [9.16] & $\ldots$ & $\ldots$ & $\ldots$ & $\ldots$ \\
    25 & 2011.80 & 2013.95 & 22.72 & 22.87 & [2031.83] & [11.50] & $\ldots$ & $\ldots$ & $\ldots$ & $\ldots$ \\
	\hline
 Means &  &  &  21.44 & 21.74 & & 10.69 & $\ldots$ & $\ldots$ & 19.67 & 19.04 \\
 Std. Dev. &  &  &  1.74 & 2.40 & & 1.27 & $\ldots$ & $\ldots$ & 1.71 & 1.88 \\
    \hline
  \end{tabular}
 \end{center}
\end{table}
\end{landscape}


\bibliographystyle{spr-mp-sola}


\end{document}